\documentclass{article} 
\usepackage{aaai2027}  
\usepackage[hyphens]{url}  
\usepackage{graphicx} 
\usepackage{natbib}  
\usepackage{caption} 
\usepackage{algorithm}
\usepackage{algorithmic}
\usepackage{amsmath}
\usepackage{booktabs}
\usepackage{multirow}
\usepackage[table]{xcolor}
\definecolor{lightgreen}{RGB}{220,245,220}
\definecolor{lightgray}{RGB}{240,240,240}
\usepackage{newfloat}
\usepackage{listings}
\DeclareCaptionStyle{ruled}{labelfont=normalfont,labelsep=colon,strut=off} 
\floatstyle{ruled}
\newfloat{listing}{tb}{lst}{}
\floatname{listing}{Listing}

\usepackage{booktabs}

\title{Towards Efficient Threat Intelligence Instruction Tuning via Quality-Aware and Optimal Transport-Guided Data Selection}

\author{
Jiyu Tian, Mingchu Li, Xiangwei Chen, Liming Chen
}
\affiliations{
Dalian University of Technology, Dalian, China 
}

\begin{document}

\maketitle

\begin{abstract}
High-quality data is essential for improving instruction tuning in threat intelligence analysis. However, existing methods for generating cybersecurity instruction data lack reliable quality assessment mechanisms, and gradient influence-based data selection methods primarily focus on single-sample matching, failing to cover diverse and complex security capabilities. To address these issues, we propose CTI-TDS, a \textbf{\underline{T}}argeted \textbf{\underline{D}}ata \textbf{\underline{S}}election method for instruction tuning in \textbf{\underline{C}}yber \textbf{\underline{T}}hreat \textbf{\underline{I}}ntelligence analysis. Specifically, we first filter high-quality CTI instructions by integrating instruction style consistency and security knowledge verification. Furthermore, we formulate gradient influence as a distribution alignment problem, learning a gradient subspace of target tasks and applying entropic regularized optimal transport to optimize distribution matching between training samples and target tasks, enabling diversity-aware data selection. We conduct extensive experiments on CTIBench across multiple large language models. Experimental results demonstrate that when fine-tuning with only a small set of high-value instruction data, CTI-TDS outperforms existing data selection methods and achieves competitive performance on threat intelligence analysis tasks. We will release the source code upon acceptance.
\end{abstract}


\section{1 Introduction}
Cyber Threat Intelligence (CTI) analysis aims to identify attack patterns, track threat behaviors, and predict potential risks from massive security data, serving as a critical component of defense systems \cite{ctisurvey,iocs, humancti}. Large language models (LLMs) have demonstrated powerful semantic understanding and knowledge reasoning capabilities, providing a new pathway for automating threat intelligence analysis \cite{cyberpalai, automaldesc, cticonnect, hypergllm, cyberpal20}. However, general-purpose LLMs lack domain-specific cybersecurity knowledge and struggle to understand complex security tasks such as vulnerability correlation analysis, attack technique identification, and threat attribution. Therefore, it is essential to equip LLMs with specialized capabilities for CTI analysis through instruction tuning \cite{primus, frp}.

Effective instruction tuning highly depends on the quality and coverage of training data \cite{lima,tulu,mates}. Existing approaches leverage large-scale domain knowledge bases and task templates to generate massive threat intelligence instructions \cite{ctibench,primus,cticonnect}. However, automatically generated data still faces two major challenges. First, the quality of generated data is difficult to guarantee. Different prompt templates may produce responses with varying linguistic styles, leading to deviations between the generated data and the expected output patterns of target models \cite{scar}. Meanwhile, generated content in the cybersecurity may suffer from issues with factual consistency, such as incorrect vulnerability descriptions and inaccurate mappings of attack techniques \cite{primus}. Furthermore, simply increasing data volume does not necessarily lead to optimal instruction tuning performance \cite{less,approxless,iprox}. CTI tasks involve diverse cybersecurity capabilities, including vulnerability analysis, CVE-CWE mapping and attack technique identification \cite{ctibench,cticonnect}. 
Thus, selecting a data subset that maximizes model capability improvement under a limited data budget remains challenging.

Recently, targeted data selection methods have improved the efficiency of instruction tuning by evaluating the influence of training samples on target tasks and selecting high-value data \cite{g2is, wici, miwv}. For example, LESS \cite{less} selects samples relevant to target tasks based on gradient influence estimation, while GIST \cite{gist} further learns the target task gradient subspace to obtain low-dimensional representations, thereby reducing computational costs and improving selection efficiency. However, these methods mainly rely on sample-level gradient similarity for ranking and only consider the matching relationship between individual samples and target tasks, making it difficult to capture the relationship between the overall training data and the capability distribution of target tasks. In the CTI scenario, selecting only samples with similar gradient directions may concentrate cybersecurity knowledge on a limited set of capability dimensions, weakening the model's ability to generalize across diverse security tasks. Therefore, effective data selection methods should jointly consider data quality and cybersecurity capability coverage.

To address the challenges above, we propose CTI-TDS, a quality-aware and optimal transport-guided data selection framework for CTI instruction tuning. Specifically, we first tackle the quality deviation issue in automatically generated CTI instructions by introducing a few target model instruction-response pairs as style constraints, while leveraging security knowledge resources to verify critical factual information in the generated responses, thereby filtering candidate data that meets the requirements of target CTI tasks. Furthermore, to address the imbalanced capability distribution, we leverage gradients from target CTI tasks to learn a task-relevant subspace, and establish a matching relationship between the candidate data distribution and the target task capability distribution via entropic regularized optimal transport, enabling us to select a training subset that exhibits both task relevance and capability coverage. We conducted extensive experiments using \textit{Qwen2.5-7B-Instruct} \cite{qwen2.5} and \textit{Llama3.1-8B-Instruct} on the CTIBench \cite{ctibench} benchmark tasks. Experimental results demonstrate that CTI-TDS can improve model performance across various CTI analysis tasks while reducing the size of training data, validating the effectiveness of quality-aware data filtering and gradient distribution alignment strategies in security-domain instruction tuning. Our contributions are as follows:

\begin{itemize}
    \item We propose a quality assessment method for security-domain instructions that integrates response style consistency and security knowledge verification to reduce stylistic deviation and factual errors in automatically generated data.
    \item We design a gradient-influence-based data selection approach that leverages low-dimensional gradient subspace modeling and entropic regularized optimal transport to enhance selection diversity while preserving task relevance.
    \item We construct CTISec-Instruct-50K, a CTI instruction tuning dataset containing 50K samples, and validate the effectiveness of CTI-TDS across multiple LLMs and CTI analysis tasks.
\end{itemize}

\section{2 Related Work}
\subsection{2.1 LLM-based Cyber Threat Intelligence Analysis}

LLMs can effectively assist various CTI analysis tasks \cite{sok,primod4ai,sigmerge,knowhow}. For example, CTIBench \cite{ctibench} first introduces an LLM-oriented CTI evaluation framework that assesses the knowledge understanding and reasoning capabilities of LLMs in cyber threat analysis through tasks such as vulnerability classification, CVSS scoring, MITRE ATT\&CK technique identification, and threat attribution. Furthermore, SoK \cite{sok} analyzes NLP and LLM-based threat technique extraction approaches and indicates that LLMs have become an important direction for automated CTI report analysis. However, existing methods remain constrained by factors such as data quality, task definition variations, and insufficient domain knowledge. To enhance the specialized capabilities of LLMs in cybersecurity, recent studies have focused on improving model adaptation to security tasks through instruction tuning. CyberPal.AI \cite{cyberpalai} constructs a security domain instruction dataset SecKnowledge based on expert knowledge and uses the dataset to train security expert LLMs. Furthermore, PRIMUS \cite{primus} investigates the data construction pipeline for cybersecurity LLMs from a comprehensive perspective, covering data resources for pre-training, instruction tuning, and reasoning stages.

However, automatically generated instructions rely on the capabilities of generation models and may suffer from inconsistent formats, objective deviations, and factual errors. Scaling up data volume alone cannot guarantee effective instruction tuning. How to automatically evaluate and select high-quality instructions remains a critical challenge for building CTI LLMs.

\subsection{2.2 Target Data Selection for Instruction Tuning}
Instruction tuning does not entirely depend on the scale of training data, but is closely related to data quality and task relevance \cite{lima, less,logra,iprox}. LESS \cite{less} first estimates the influence of each sample on the target task and selects representative samples from large-scale instructions. Approx-LESS \cite{approxless} reduces computational cost by training a regression model to predict influence scores. To further reduce computational overhead, IPROX \cite{iprox} introduces an influence-preserving proxy model directly derived from the target model, enabling small-scale models to inherit the influence characteristics of the target model while maintaining computational efficiency. Beyond efficiency, BIDS \cite{bids} points out that selecting data solely based on influence scores tends to introduce task bias, causing the selected samples to neglect other capability dimensions. G2IS \cite{g2is} emphasizes the joint structural relationships among training samples, arguing that independently scoring individual samples fails to yield a globally optimal subset. T-SHIRT \cite{tshirt} analyzes the limitations of traditional scoring functions from the perspective of sample-wise local consistency. GIST \cite{gist} further improves gradient selection from an optimization geometry perspective, constructing a task-specific low-dimensional subspace from validation gradients.

However, these methods still frame data selection as a sample-level matching problem, in which the relevance between each training sample and the target task is calculated independently, failing to capture the relationship between the distribution of candidate instructions and the distribution of target task capabilities.

\begin{figure*}
\centering
\includegraphics[width=0.98\linewidth]{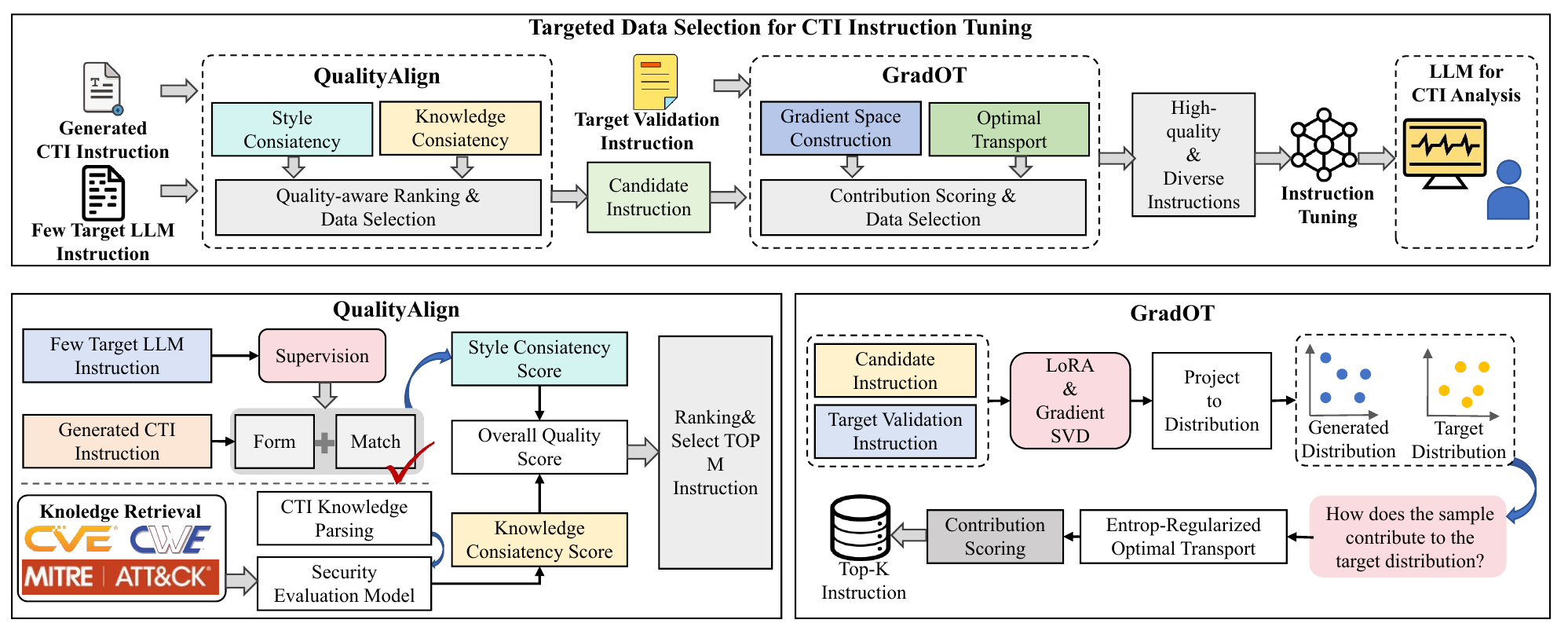}
\caption{Overview of the CTI-TDS approach} 
\label{framework}
\end{figure*}

\section{3 Problem Formulation}
Given a large-scale security instruction dataset
$\mathcal{D}=\{z_i\}_{i=1}^{N}$, where each sample is represented as
$z_i=(x_i,y_i)$, $x_i$ denotes the user instruction and $y_i$ denotes the target response.
For CTI tasks, $x_i$ represents a security query and $y_i$ represents the corresponding analysis result. Targeted data selection is to select a subset
$\mathcal{S}\subset\mathcal{D}$ with size $K$, such that the selected data can effectively improve model performance on target CTI tasks.
Since CTI analysis involves multiple cybersecurity capabilities, an ideal data selection strategy should ensure that the selected data covers diverse security capability dimensions.
Therefore, we formulate targeted data selection as:

\begin{equation}
	\mathcal{S}^{*}
	=
	\arg\max_{\mathcal{S}\subset\mathcal{D},|\mathcal{S}|=K}
	\sum_{z_i\in\mathcal{S}}
	\mathcal{I}(z_i,\mathcal{D}_{val})
	+
	\lambda
	\Omega(\mathcal{S},\mathcal{C})
\end{equation}

where $\mathcal{I}(z_i,\mathcal{D}_{val})$ denotes the influence of sample $z_i$ on the target CTI task, $\Omega(\mathcal{S},\mathcal{C})$ measures the coverage of the selected data over the CTI capability space,
and $\lambda$ balances the importance between sample influence and capability coverage.
$\mathcal{C}=\{c_1,c_2,\dots,c_m\}$ denotes the CTI capability space.
Each $c_j$ represents a specific security analysis capability.

Our goal is to design a targeted data selection method for CTI analysis that identifies instructions with high task relevance and capability coverage under a limited data budget, thereby improving the analysis performance of LLMs.

\section{4 Methodology}
\subsection{4.1 Overview}
To address the challenges of consistency and diversity in instruction tuning, we propose a targeted data selection framework for CTI, comprising two modules: QualityAlign and GradOT. QualityAlign jointly evaluates response style consistency and security knowledge consistency to construct a reliable candidate dataset $\mathcal{D}{q}$. GradOT then selects a subset $\mathcal{D}{k}$ with strong task relevance and broad capability coverage through gradient distribution alignment.

\textbf{QualityAlign} addresses style mismatch and unreliable security knowledge in generated data. Inspired by instruction quality evaluation \cite{scar} and security-domain data construction practices \cite{primus}, it introduces a small set of high-quality reference instruction-response pairs as style constraints, and leverages security knowledge resources such as CVE, CWE, and MITRE ATT\&CK to verify key factual content in generated responses. By jointly assessing response structure, task alignment, and security knowledge consistency, QualityAlign filters samples that satisfy target CTI requirements.

Given the candidate dataset $\mathcal{D}{q}$, \textbf{GradOT} overcomes the limitation of sample-level gradient matching by considering the overall capability distribution. By constructing a low-dimensional task representation space based on the gradients of the target CTI task, GradOT represents candidate and target tasks as probability distributions within this gradient space and learns the matching between them via optimal transport. Based on sample contributions to the target capability distribution, GradOT selects data $\mathcal{D}{k}$ with both task relevance and capability coverage for instruction tuning and CTI evaluation.

\subsection{4.2 Quality Alignment for Instruction Curation}
Generated CTI instructions expand cybersecurity training resources but often contain style deviations and security knowledge errors. To address these issues, QualityAlign jointly evaluates response style and security knowledge consistency to construct a reliable candidate dataset $\mathcal{D}_{q}$.

\subsubsection{Style Consistency Evaluation}
Maintaining consistency with the expected expression patterns of the target LLM can reduce generation bias and improve instruction tuning performance \cite{scar}. Therefore, QualityAlign uses a few high-quality reference instruction-response pairs from target LLM as output style constraints and measures the consistency between candidate samples and reference data. Given a candidate instruction-response pair $z_i=(x_i,y_i)$ and a reference set
$\mathcal{D}_{l}=\{(x_j^r,y_j^r)\}_{j=1}^{L}$,
we evaluate style consistency from two dimensions: output-format consistency and task-response consistency.

\begin{equation}
	\begin{aligned}
		\mathcal{P}_{style}(z_i)
		=&
		\max_{z_j^r\in\mathcal{D}_{l}}
		\operatorname{sim}
		(h(y_i),h(y_j^r))\\
		&+
		\max_{z_j^r\in\mathcal{D}_{l}}
		\operatorname{sim}
		(h(x_i,y_i),h(x_j^r,y_j^r))
	\end{aligned}
\end{equation}

where the first term denotes output-format consistency $\mathcal{P}_{\mathrm{form}}$, which measures similarities in structural organization, analysis procedures, and professional expression between candidate and reference responses. The second term denotes task-response consistency $\mathcal{P}_{\mathrm{match}}$, which measures consistency in task understanding and response objectives between candidate and reference instruction-response pairs. Here, following prior work on LLM-based text representations \cite{htp}, $h(\cdot)$ is implemented by mean-pooling the final-layer hidden states of the frozen target LLM over non-padding tokens, followed by $L_2$ normalization. $\operatorname{sim}(\cdot,\cdot)$ denotes the similarity function.

\subsubsection{Knowledge Consistency Evaluation}
Style consistency alone cannot ensure the factual reliability of generated CTI instructions. Therefore, QualityAlign further evaluates security knowledge consistency using external knowledge resources, combining knowledge retrieval with a security-aware evaluation model to verify critical facts in generated responses. 
For each candidate response $y_i$, QualityAlign first identifies relevant security concepts and entities, such as vulnerability identifiers, vulnerability types, attack techniques, threat behaviors, and mitigation information. Based on the identified concepts, relevant evidence $\mathcal{E}_i$ is retrieved from external security resources including CVE, CWE, and MITRE ATT\&CK. Thus, the knowledge consistency score is computed as:

\begin{equation}
\mathcal{P}_\mathrm{Know}(z_i)=J_{\psi}(x_i,y_i,\mathcal{E}_i)
\end{equation}

where $\mathcal{E}_i$ denotes the retrieved CTI evidence associated with candidate
response $y_i$. $J_{\psi}(\cdot)$ denotes the LLM-based consistency
evaluator \cite{primus} with an output range of $[0,1]$, which measures whether
the security claims in the response are consistent with the retrieved
evidence. A higher $P_{\mathrm{Know}}(z_i)$ indicates that the response
contains more evidence-supported CTI knowledge.

\subsubsection{Quality-aware Candidate Construction}
Based on response style consistency and security knowledge consistency evaluation, QualityAlign integrates style and security knowledge consistency to rank generated CTI instructions and construct a reliable candidate pool. For each sample $z_i$, the overall quality score is defined as:

\begin{equation}
	Q(z_i)=\beta \mathcal{P}_{style}(z_i)+(1-\beta)\mathcal{P}_{Know}(z_i)
\end{equation}

where $\beta$ controls the relative contribution of the two consistency scores. A higher $Q(z_i)$ indicates stronger consistency in output format, task response, and security facts. QualityAlign retains the top-$M$ samples from the synthetic dataset $\mathcal{D}$ to construct the candidate dataset $\mathcal{D}_{q}$, reducing style deviations and factual errors, providing reliable candidates for GradOT to further optimize task relevance and capability coverage.

\subsection{4.3 Gradient-guided Optimal Transport for Targeted Data Selection}
Although QualityAlign improves CTI instruction quality, the candidate dataset may still contain capability imbalance and redundant samples. Therefore, GradOT models samples in the gradient space and aligns their distribution with target CTI tasks through optimal transport, selecting a subset from $\mathcal{D}_{q}$ that preserves capability coverage while reducing redundancy.

\subsubsection{Gradient Space Construction}
To estimate how candidate samples influence the optimization of target CTI tasks, GradOT constructs gradient representations in the parameter-efficient fine-tuning space. Using LoRA, the gradient of each sample $z_i\in\mathcal{D}_{q}$ is computed as:

\begin{equation}
	g_i=\nabla_{\theta_L}\ell(z_i;\theta_L)
\end{equation}

where $\theta_L$ denotes the trainable LoRA parameters, $\ell(\cdot)$ is the instruction tuning loss, and $g_i$ characterizes the contribution of $z_i$ to model updates. Directly comparing high-dimensional gradients is computationally expensive and may retain directions irrelevant to the target tasks. Inspired by \cite{gist}, GradOT therefore learns a task-relevant gradient subspace from the target validation set $\mathcal{D}_{val}$. Specifically, we construct the validation gradient matrix $G_{val}$ and apply singular value decomposition:

\begin{equation}
	G_{val}=U\Sigma V^{T},
	\quad
	U_{c}=[u_1,u_2,\dots,u_r]
\end{equation}

where $U_{c}$ denotes the target CTI gradient subspace formed by the left singular vectors corresponding to the top-$r$ largest singular values. Then, candidate sample gradients are projected into the subspace using $\hat g_i=U_c^Tg_i$, and the projected gradients are directly used as latent gradient representations $r_i$ for subsequent transport alignment. Through this process, GradOT transforms original high-dimensional gradients into target task-relevant representations, preserves the major optimization directions related to CTI capabilities, and filters irrelevant gradient noise, providing a unified representation space for gradient distribution alignment in subsequent optimal transport optimization.

\subsubsection{Optimal Transport-based Distribution Alignment}
Existing gradient-guided methods rank samples by gradient similarity to target tasks, but overly rely on local relevance, leading to redundant selection of similar-gradient samples and insufficient coverage of diverse CTI capabilities. To address this, GradOT models the gradient representations of candidate data and target tasks as empirical distributions and learns their global alignment via semi-relaxed optimal transport. Specifically, the candidate and target distributions are represented as:

\begin{equation}
P=\sum_i a_i\delta_{r_i},
Q=\sum_j b_j\delta_{r_j^{val}},
C_{ij}=d(r_i,r_j^{val})
\end{equation}

where $a_i$ and $b_j$ denote the distribution weights of candidate samples and target samples, respectively, $\delta$ denotes the Dirac measure, and $C_{ij}$ denotes the transport cost between the latent gradient representations of candidate sample $i$ and target sample $j$. A smaller $C_{ij}$ indicates that the two samples have more similar task optimization directions, meaning that the candidate sample can better support the learning of target CTI capabilities. GradOT then uses entropy-regularized optimal transport to solve the optimal transport plan between candidate samples and the target task distribution:

\begin{equation}
\pi^*=\arg\min_{\pi\ge0}\langle \pi,C\rangle-\epsilon H(\pi)
\end{equation}

where $\pi$ denotes the transport plan between the candidate data distribution and the target task distribution, $\epsilon$ denotes the entropy regularization coefficient, and $H(\pi)$ denotes the entropy regularization term used to obtain a smooth and stable transport relationship. By optimizing the transport plan, GradOT measures the contribution of candidate data to the target-task gradient distribution from a global perspective rather than relying only on local similarity between individual samples and the target task. Different from balanced OT, we only constrain the target marginal, allowing candidate examples to receive different transport masses.

\subsubsection{Targeted Data Selection}

Based on the optimal transport matrix $\pi^{*}$, GradOT further measures the contribution of each sample to the target CTI task distribution according to its transport mass$M_i=\sum_j\pi_{ij}^{*}$, where $\mathrm{M}_i$ represents the contribution of candidate sample $z_i$ to the target CTI task distribution. A larger $\mathrm{M}_i$ indicates that the sample can cover more target task-relevant gradient directions and therefore has a higher data selection value. Finally, we rank samples according to their contribution scores and select the final training set under the budget constraint $\mathrm{K}$:

\begin{equation}
	\mathcal{S}^{*}=\mathrm{Top} \mathrm{K} (M_i,\mathcal{D}_{q})
\end{equation}

Unlike gradient similarity methods, GradOT aligns candidate samples with the target-task gradient distribution via optimal transport, reducing redundancy while preserving diverse optimization directions.

\section{5 Experiments}

\begin{table*}[t]
\centering
\setlength{\tabcolsep}{4pt}
\footnotesize
\begin{tabular}{llccccc}
\toprule
\textbf{Model} & \textbf{Method} & \textbf{Data Size} 
& \textbf{CTI-MCQ (Acc$\uparrow$)} & \textbf{CTI-RCM (Acc$\uparrow$)} & \textbf{CTI-VSP (MAD$\downarrow$)} & \textbf{CTI-ATE (F1$\uparrow$)} \\
\midrule
\multirow{8}{*}{Qwen2.5-7B-Instruct}
& Base & / & 63.28\% & 55.79\% & 1.6970 & 0.1995 \\
\rowcolor{lightgray}
& Full Data & 50K & 65.52\% & 58.24\% & 1.5907 & 0.2324 \\
& Random & 2.5K & 53.96\% & 49.40\% & 1.9370 & 0.1118 \\
& LESS & 2.5K & 64.40\% & 57.33\% & 1.6191 & 0.2208 \\
& SCAR & 2.5K & 62.11\% & 56.48\% & 1.7482 & 0.2103 \\
& GIST & 2.5K & 64.78\% & 57.91\% & 1.5825 & 0.2268 \\
& PRIMUS-Instruct & <1K & 62.16\% & 56.90\% & 1.7723 & 0.1155 \\
\rowcolor{lightgreen}
& \textbf{CTI-TDS (Ours)} & 2.5K 
& \textbf{66.24\%} & \textbf{59.33\%} 
& \textbf{1.5238} & \textbf{0.2417} \\
\midrule

\multirow{8}{*}{Llama-3.1-8B-Instruct}
& Base & / & 53.96\% & 57.00\% & 1.8617 & 0.1763 \\
\rowcolor{lightgray}
& Full Data & 50K & 57.51\% & 60.23\% & 1.7031 & 0.2151 \\
& Random & 2.5K & 16.76\% & 44.84\% & 2.3104 & 0.0912 \\
& LESS & 2.5K & 56.64\% & 58.12\% & 1.7728 & 0.2064 \\
& SCAR & 2.5K & 52.88\% & 56.13\% & 1.9017 & 0.1882 \\
& GIST & 2.5K & 58.84\% & 59.16\% & 1.7241 & 0.2157 \\
& PRIMUS-Instruct & <1K & 42.32\% & 55.90\% & 1.7825 & 0.0939 \\
\rowcolor{lightgreen}
& \textbf{CTI-TDS (Ours)} & 2.5K 
& \textbf{59.23\%} & \textbf{61.44\%} 
& \textbf{1.6472} & \textbf{0.2284} \\
\bottomrule
\end{tabular}
\caption{Main comparison of different selection methods on CTI instruction tuning. Best results are highlighted in light green.}
\label{tab:main_comparison}
\end{table*}

\subsection{5.1 Datasets}
\textbf{Instruction Tuning:} We construct CTISec-Instruct-50K, a dataset comprising 50,000 CTI instruction-response pairs. We first collect 3,406 verifiable security evidence entries from CVE, CWE, and MITRE ATT\&CK, covering tasks such as vulnerability analysis, threat intelligence analysis, and attack technique identification. We then use the DeepSeek-V4-Flash API \cite{deepseekv4} to simulate security analysts, generating instruction-response pairs based on the evidence, and organize the content according to predefined templates for analysis, impact, detection, and mitigation. The dataset contains no invalid JSON or duplicate records and can be directly used for supervised fine-tuning of LLMs. Generation details are provided in Appendix A.

\textbf{Evaluation Data:} We assess the effectiveness of each method on CTIBench \cite{ctibench}, a benchmark that includes multiple tasks requiring security knowledge understanding and reasoning. Following existing work \cite{primus}, we select four core tasks: CTI-MCQ with 2,500 samples for evaluating cybersecurity knowledge comprehension; CTI-RCM with 1,000 samples for assessing CVE-to-CWE root cause mapping; CTI-VSP with 1,000 samples for CVSS v3.1 vector prediction; and CTI-ATE with 60 samples and 397 ATT\&CK technique labels for attack technique identification. These tasks cover key capability dimensions including knowledge comprehension, vulnerability analysis, risk assessment, and attack behavior recognition, providing a comprehensive measure of model adaptation after security instruction tuning. Full details are provided in Appendix A.

\subsection{5.2 Evaluation Metrics}
We follow the CTIBench \cite{ctibench} evaluation protocol. CTI-MCQ and CTI-RCM are classification tasks, and we adopt accuracy as the evaluation metric. For CTI-VSP, the model is required to predict the CVSS vector, which we convert to CVSS scores and compute the Mean Absolute Deviation from the ground truth scores. A lower MAD indicates better prediction accuracy. CTI-ATE is a multi-label classification task, and we use Micro-F1 to measure the model's comprehensive ability to identify MITRE ATT\&CK techniques from threat reports. These metrics evaluate the model's threat intelligence analysis capability from four dimensions: knowledge comprehension, root cause mapping, threat scoring, and attack identification.

\subsection{5.3 Baselines}
We select Qwen2.5-7B-Instruct \cite{qwen2.5} and Llama-3.1-8B-Instruct as the base models and compare against the following baselines: Base refers to the base model without fine-tuning; Full Data uses all 50K training samples; Random randomly selects samples from the candidate set; LESS \cite{less} selects high-value data by estimating the influence of each training sample on the target task; SCAR \cite{scar} serves as a representative of general instruction quality filtering by evaluating response style consistency; GIST \cite{gist} further reduces noise using the target task gradient subspace based on LESS; PRIMUS-Instruct \cite{primus} represents security-domain instruction data construction. 

\subsection{5.4 Experimental Environment}

All experiments are conducted on a workstation equipped with 8 NVIDIA L40 GPUs, each with 48GB memory, an Intel Xeon Gold 6326 CPU, and 503GB RAM. The software environment includes Ubuntu 22.04, CUDA 12.6, Python 3.9.21, and PyTorch 2.6.0. Batch size is set to 128, the learning rate to $2\times 10^{-5}$, and we train for 2 epochs. Gradient checkpointing and BF16 mixed precision are used to reduce memory consumption. Full fine-tuning takes approximately 6 hours for Qwen2.5-7B-Instruct and 7.5 hours for Llama-3.1-8B-Instruct. Details are provided in Appendix B.

\subsection{5.5 Main Comparison}

To evaluate CTI-TDS, we conduct experiments on four tasks, including CTI-MCQ, CTI-RCM, CTI-VSP, and CTI-ATE, using Qwen2.5-7B-Instruct and Llama-3.1-8B-Instruct. As shown in Table~\ref{tab:main_comparison}, CTI-TDS achieves the best performance across all tasks and both models.

On Qwen, Full Data training with 50K samples reaches 65.52\% MCQ accuracy and 0.2324 ATE F1, while Random selection achieves only 53.96\% and 0.1118, showing that unguided sampling introduces substantial noise. Gradient-based methods perform better. LESS obtains 64.40\% CTI-MCQ accuracy and 0.2208 CTI-ATE F1, while GIST further improves them to 64.78\% and 0.2268 through gradient subspace modeling. SCAR achieves 62.11\% CTI-MCQ accuracy and 0.2103 CTI-ATE F1, suggesting that style-based filtering alone lacks sufficient task awareness. PRIMUS-Instruct uses fewer than 1K curated samples but achieves only 0.1155 CTI-ATE F1, indicating limited coverage of complex CTI capabilities. In contrast, CTI-TDS achieves 66.24\% MCQ accuracy, 59.33\% RCM accuracy, 1.5238 CTI-VSP MAD, and 0.2417 CTI-ATE F1 with only 2.5K samples.

Similar trends are observed on Llama. CTI-TDS reaches 59.23\% MCQ accuracy, 61.44\% RCM accuracy, 1.6472 CTI-VSP MAD, and 0.2284 ATE F1, outperforming both Full Data and all selection baselines. In particular, it exceeds Full Data by 1.72 percentage points on CTI-MCQ, 1.21 percentage points on CTI-RCM, and 0.0133 on CTI-ATE F1. These results demonstrate that jointly optimizing quality-aware filtering and gradient distribution alignment enables CTI-TDS to reduce redundant or unreliable samples while preserving task relevance and capability coverage. The gains on CTI-ATE and CTI-VSP further confirm the importance of high-quality and diverse data for reasoning-intensive threat intelligence analysis.

\begin{table*}[t]
\centering
\setlength{\tabcolsep}{7pt}
\footnotesize
\begin{tabular}{lcccccccc}
\toprule
\multirow{2}{*}{\textbf{Components}} 
& \multicolumn{4}{c}{\textbf{Qwen2.5-7B-Instruct}}
& \multicolumn{4}{c}{\textbf{Llama-3.1-8B-Instruct}} \\
\cmidrule(lr){2-5}
\cmidrule(lr){6-9}
& \textbf{CTI-MCQ} 
& \textbf{CTI-RCM} 
& \textbf{CTI-VSP} 
& \textbf{CTI-ATE}
& \textbf{CTI-MCQ} 
& \textbf{CTI-RCM} 
& \textbf{CTI-VSP} 
& \textbf{CTI-ATE} \\
\midrule

Baseline
& 62.16\% & 56.90\% & 1.7723 & 0.1155
& 42.32\% & 55.90\% & 1.7825 & 0.0939 \\

w/o QualityAlign
& 63.85\% & 56.81\% & 1.6371 & 0.2235
& 53.10\% & 56.30\% & 1.8800 & 0.1900 \\

w/o GradOT
& 65.20\% & 58.30\% & 1.5600 & 0.2310
& 57.89\% & 59.87\% & 1.6994 & 0.2177 \\

\textbf{Proposed}
& \textbf{66.24\%} & \textbf{59.33\%} & \textbf{1.5238} & \textbf{0.2417}
& \textbf{59.23\%} & \textbf{61.44\%} & \textbf{1.6472} & \textbf{0.2284} \\
\bottomrule
\end{tabular}
\caption{Ablation study of QualityAlign and GradOT on CTI instruction tuning.}
\label{tab:ablation}
\end{table*}

\subsection{5.6 Sensitivity Analysis}

\begin{figure}[t]
    \centering
    \includegraphics[width=0.49\linewidth]{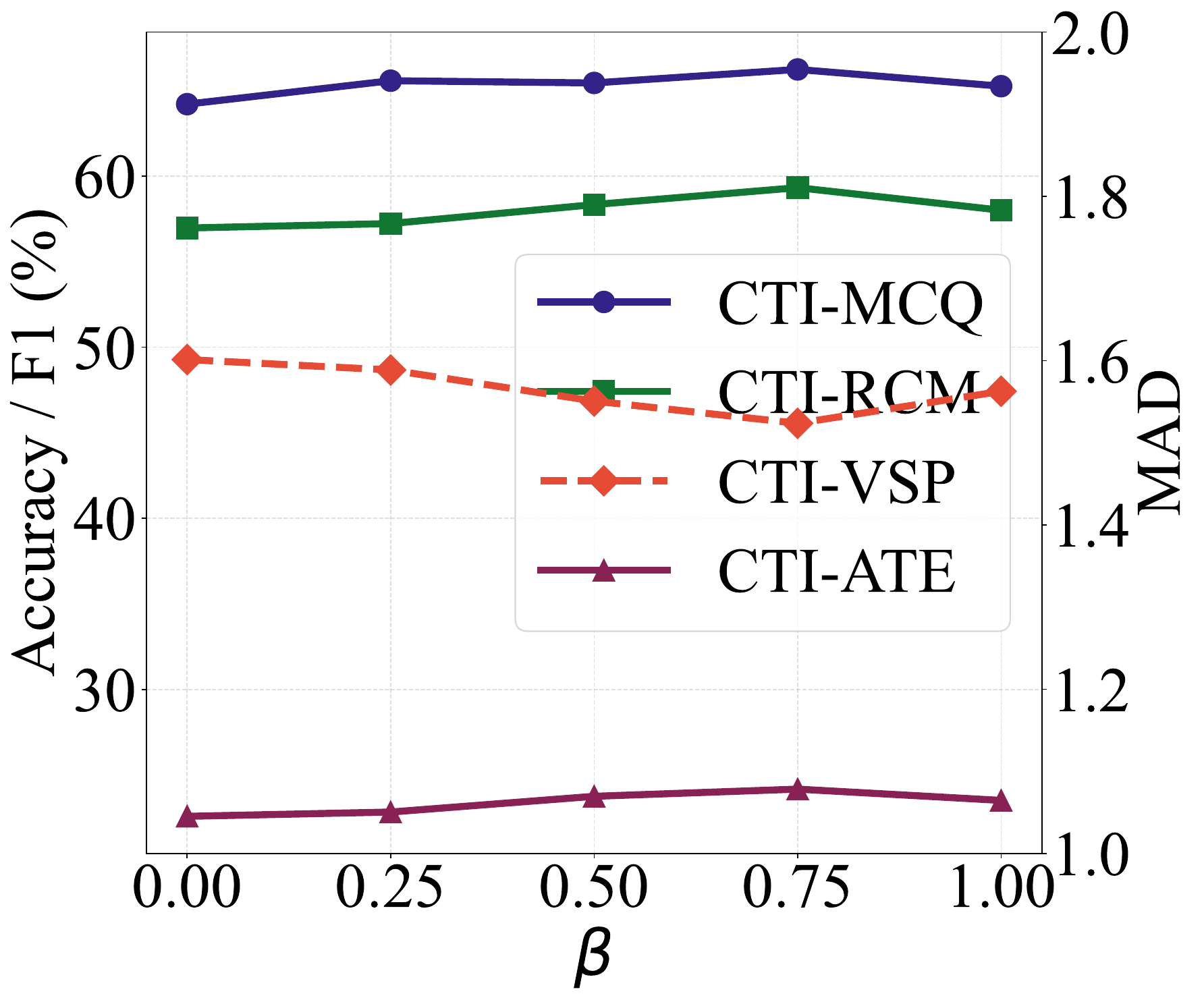}
    \hfill
    \includegraphics[width=0.49\linewidth]{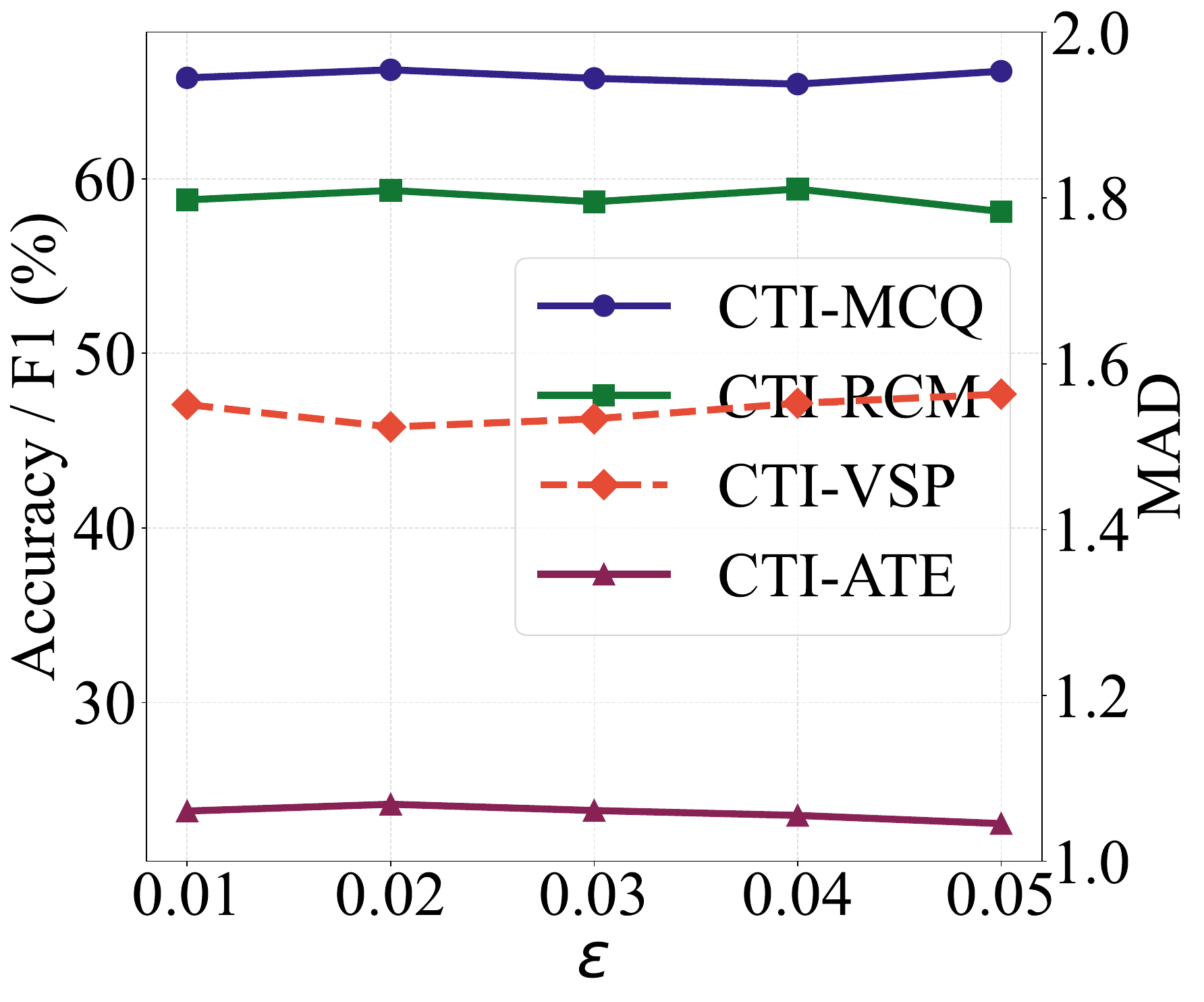}
    \caption{
    Sensitivity analysis of CTI-TDS with different hyperparameters.
    The left figure evaluates the influence of the QualityAlign weight $\beta$,
    and the right figure evaluates the influence of the GradOT entropy regularization coefficient $\epsilon$.
    }
    \label{fig:sensitivity_analysis}
\end{figure}

We analyze the sensitivity of two key hyperparameters, $\beta$ and $\epsilon$, by varying one parameter while keeping the remaining configurations fixed. As shown in Figure~\ref{fig:sensitivity_analysis}, CTI-TDS remains stable across a broad range of settings. We provide more details in Appendix C.

When $\beta$ varies from 0 to 1.0, the method maintains competitive performance on all four tasks. At $\beta=0$, QualityAlign relies only on security knowledge consistency and achieves 64.21\% on CTI-MCQ, 56.97\% on CTI-RCM, 1.6014 MAD on CTI-VSP, and 0.2258 F1 on CTI-ATE. Incorporating style consistency gradually improves the overall results, with the best performance obtained at $\beta=0.75$, reaching 66.24\%, 59.33\%, 1.5238, and 0.2417, respectively. Increasing $\beta$ to 1.0 removes security knowledge verification and causes a moderate decline, although the results remain better than those at $\beta=0$.

CTI-TDS is also robust to $\epsilon$ within the range of 0.01 to 0.05. The best configuration is $\epsilon=0.02$, which achieves 66.24\% on CTI-MCQ, 59.33\% on CTI-RCM, 1.5238 MAD on CTI-VSP, and 0.2417 F1 on CTI-ATE. Deviating from this value introduces only limited fluctuations. At $\epsilon=0.01$, CTI-MCQ remains at 65.78\%, while CTI-VSP MAD increases by only 0.0271. Even at $\epsilon=0.05$, the method obtains 66.16\% on CTI-MCQ, 1.5634 MAD on CTI-VSP, and 0.2306 F1 on CTI-ATE. These results indicate that GradOT remains effective under different entropy regularization strengths.

Overall, CTI-TDS shows stable performance across both hyperparameters. We therefore use $\beta=0.75$ and $\epsilon=0.02$ as the default configuration in subsequent experiments.

\subsection{5.7 Data Efficiency Analysis}

\begin{figure}[t]
    \centering
    \includegraphics[width=0.49\linewidth]{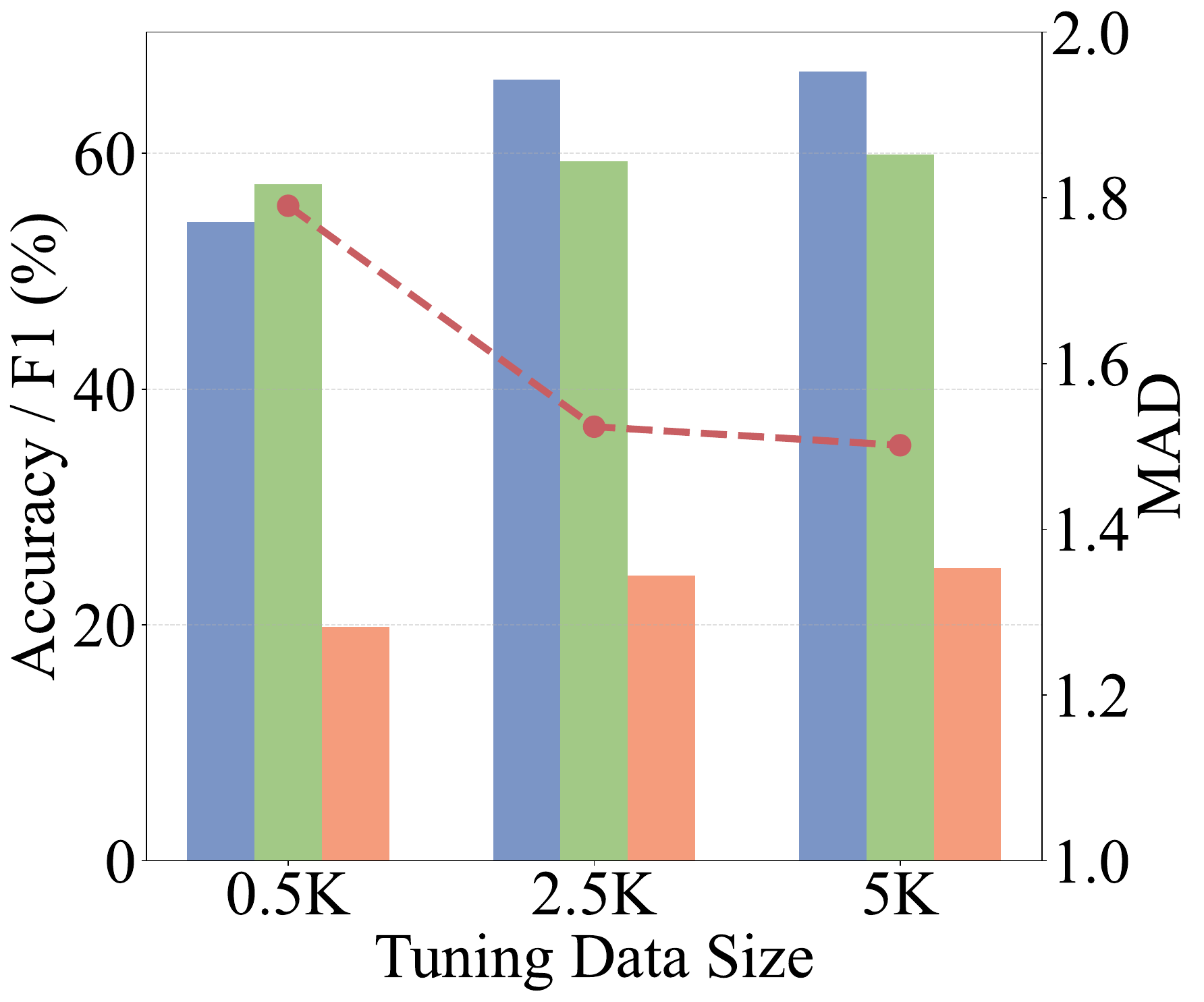}
    \hfill
    \includegraphics[width=0.49\linewidth]{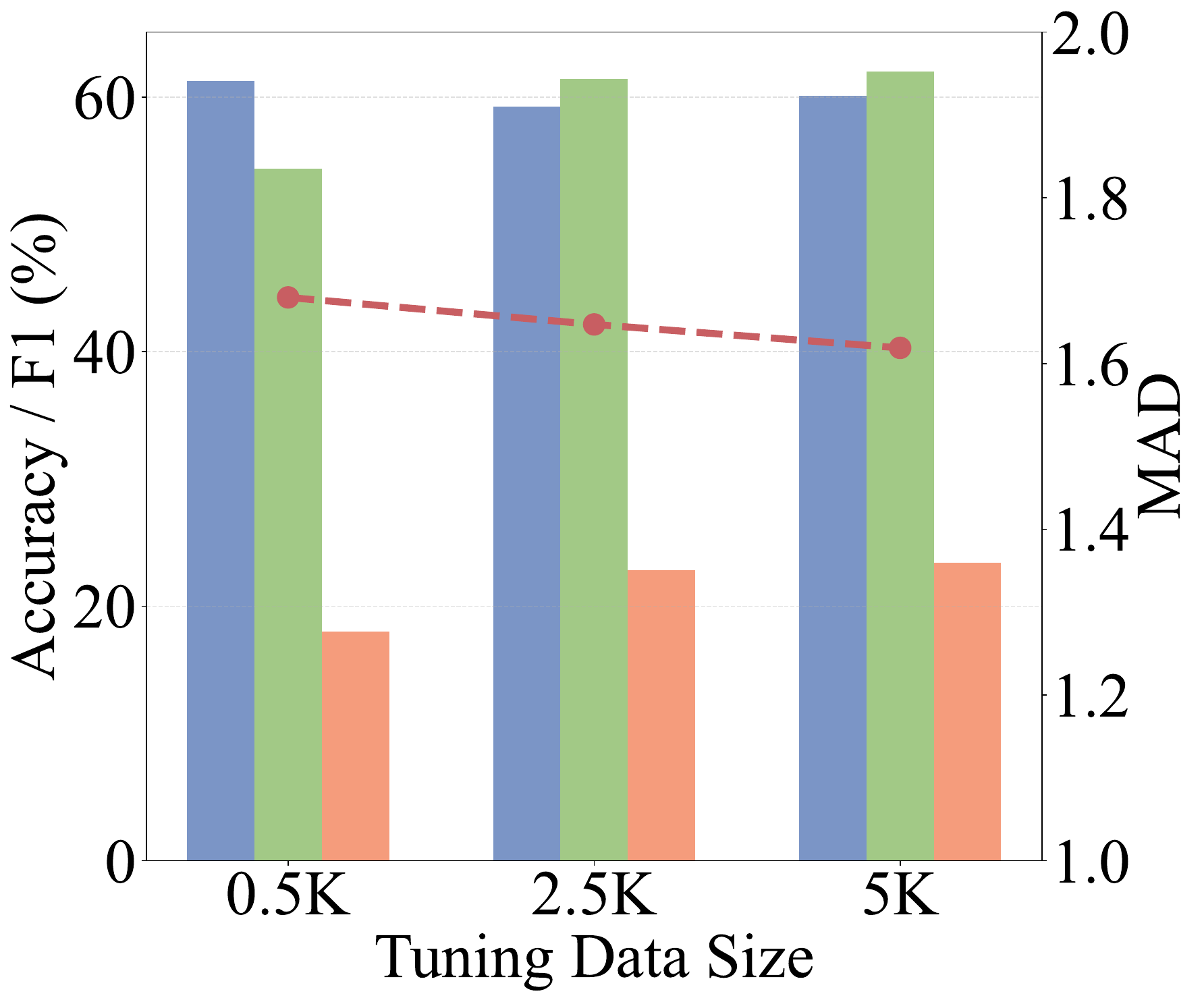}
    
    \includegraphics[width=0.6\linewidth]{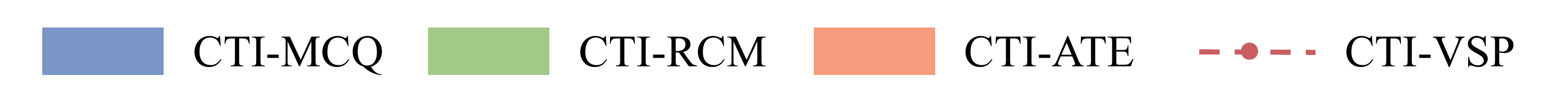}
    \caption{
Data efficiency of CTI-TDS on Qwen2.5-7B-Instruct and Llama-3.1-8B-Instruct under different selection ratios.
    }
    \label{fig:data_efficiency}
\end{figure}

We evaluate the data efficiency of CTI-TDS by increasing the selection ratio from 1\% to 10\% and measuring performance on four CTI tasks. As shown in Figure~\ref{fig:data_efficiency}, most gains are obtained when the data budget increases from 0.5K to 2.5K, while further expansion to 5K provides smaller improvements. See Appendix C for more details.

On Qwen2.5-7B-Instruct, increasing the data size from 0.5K to 2.5K improves CTI-MCQ from 54.16\% to 66.24\% and CTI-RCM from 57.38\% to 59.33\%. Meanwhile, CTI-VSP MAD decreases from 1.7904 to 1.5238, and CTI-ATE F1 increases from 0.1982 to 0.2417. Expanding the dataset to 5K yields only marginal gains, reaching 66.94\%, 59.92\%, 1.5014, and 0.2483 on the four tasks, respectively. Despite using only 10\% of the original data, the 5K setting outperforms Full Data with 50K samples on all metrics, demonstrating the effectiveness of selecting high-value samples.

On Llama-3.1-8B-Instruct, CTI-MCQ shows a non-monotonic trend, with 61.28\% at 0.5K, 59.23\% at 2.5K, and 60.08\% at 5K. This may result from a higher concentration of MCQ-relevant samples in the smallest subset. In contrast, the other tasks consistently benefit from larger budgets. CTI-RCM increases from 54.39\% to 62.01\%, CTI-VSP MAD decreases from 1.6798 to 1.6189, and CTI-ATE F1 improves from 0.1802 to 0.2341. At 5K, CTI-TDS surpasses Full Data on CTI-RCM and CTI-ATE, achieving 60.23\% and 0.2151, respectively. By combining quality-aware filtering with gradient distribution alignment, CTI-TDS effectively covers the target-task gradient distribution under a small data budget.


\subsection{5.8 Ablation Study}
Using PRIMUS-Instruct as a baseline, we evaluate the contributions of QualityAlign and GradOT by removing each module individually. As shown in Table~\ref{tab:ablation}, the complete method consistently achieves the best performance across all tasks and both models.

On Qwen2.5-7B-Instruct, removing QualityAlign yields 63.85\% on CTI-MCQ, 56.81\% on CTI-RCM, 1.6371 MAD on CTI-VSP, and 0.2235 F1 on CTI-ATE. Removing GradOT achieves stronger results of 65.20\%, 58.30\%, 1.5600, and 0.2310, respectively, indicating the importance of quality-aware filtering. The complete method further improves these results to 66.24\%, 59.33\%, 1.5238, and 0.2417, confirming the additional benefit of gradient distribution alignment.

Similar trends are observed on Llama-3.1-8B-Instruct. The configuration without QualityAlign obtains 53.10\% CTI-MCQ accuracy and 0.1900 CTI-ATE F1, while removing GradOT reaches 57.89\% and 0.2177. The complete method achieves 59.23\% CTI-MCQ accuracy, 61.44\% CTI-RCM accuracy, 1.6472 CTI-VSP MAD, and 0.2284 CTI-ATE F1. Notably, GradOT alone produces a CTI-VSP MAD of 1.8800, whereas the complete method reduces it to 1.6472.

Overall, QualityAlign improves candidate quality through style and security knowledge consistency, while GradOT enhances task relevance and capability coverage through optimal transport. Removing either module degrades performance, demonstrating their complementary effects and the necessity of joint optimization.

\section{6 Conclusion}

In this paper, we propose CTI-TDS, an instruction tuning data selection framework for threat intelligence analysis, which aims to address the consistency and diversity challenges in instruction tuning data. We first jointly model response style consistency and security knowledge consistency to filter reliable CTI instruction samples, reducing style bias and factual errors in generated data. Then, we formulate data selection as a distribution alignment problem in the gradient space and achieve better task relevance and distribution coverage through target gradient subspace learning and entropy-regularized optimal transport. Experimental results on CTIBench demonstrate that CTI-TDS achieves superior CTI analysis performance compared with existing methods while significantly reducing the training data scale. In the future, we will further explore continual data selection methods for dynamic threat environments and improve data utilization efficiency of cybersecurity LLMs by incorporating more fine-grained security knowledge representations.



\bibliography{aaai2027}


\end{document}